# Epitaxial growth and magnetoelectric relaxor behavior in multiferroic $0.8Pb(Fe_{1/2}Nb_{1/2})O_3$-$0.2Pb(Mg_{1/2}W_{1/2})O_3$ thin films

Wei Peng,[1,a] N. Lemée,[1,b] J.-L. Dellis,[1] V. V. Shvartsman,[2] P. Borisov,[2] W. Kleemann,[2] Z. Trontelj,[3] J. Holc,[4] M. Kosec,[4] R. Blinc,[4] and M. G. Karkut[1]

[1]LPMC, Université de Picardie Jules Verne, 33 rue Saint Leu, 80039 Amiens, France

[2]Angewandte Physik, Universität Duisburg-Essen, D-47048 Duisburg, Germany

[3]Institute of Mathematics, Physics and Mechanics, 1000 Ljubljana, Slovenia

[4]Jožef Stefan Institute, Jamova 39, 1000 Ljubljana, Slovenia

[a] wei.peng@u-picardie.fr

[b] nathalie.lemee@sc.u-picardie.fr

**ABSTRACT**

We present electric and magnetic properties of $0.8Pb(Fe_{1/2}Nb_{1/2})O_3-0.2Pb(Mg_{1/2}W_{1/2})O_3$ films epitaxially grown on (001) $SrTiO_3$ substrates using pulsed laser deposition. A narrow deposition window around 710 $^oC$ and 0.2 mbar has been identified to achieve epitaxial single-phase thin films. A typical Vogel-Fulcher relaxorlike dielectric and magnetic susceptibility dispersion is observed, suggesting magnetoelectric relaxor behavior in these films similar to the bulk. We determine a magnetic cluster freezing temperature of 36 K, while observing weak ferromagnetism via magnetic hysteresis loops up to 300 K.

Recent investigations on the multiferroic ceramic solid solution $0.8Pb(Fe_{1/2}Nb_{1/2})O_3-0.2Pb(Mg_{1/2}W_{1/2})O_3$ (0.8PFN-0.2PMW) have revealed intriguing properties: both electric and magnetic relaxor behavior is reported as well as a magnetoelectric coupling effect.[1, 2] In this pseudobinary system, PFN is a *B*-site disordered compound whereas PMW is ordered. The addition of 20 % of the antiferroelectric PMW to the relaxor ferroelectric PFN compound suppresses the paraelectric to ferroelectric phase transition observed for PFN around 385 K. PMW is diamagnetic[3], and it also suppresses the long range antiferromagnetic order observed in PFN below a Néel temperature $T_N$ of ~ 145 K.[4, 5] Thus 0.8PFN-0.2PMW is site and charge disordered whose structure is cubic ($a \approx 4.014$ Å) at room temperature (RT).[6] It shows a typical relaxorlike dielectric dispersion (~ 250 K) as well as relaxor-like, or spin glass, magnetic susceptibility dispersive peaks (~ 24 K), suggesting both magnetic and electric mesoscopic orders rather than the usual long-range orders studied in multiferroic compounds. Moreover, a magnetoelectric effect was demonstrated by a significant magnetic anomaly measured at the dielectric Vogel-Fulcher (VF) freezing temperature ($T_{VF}$ ~ 245 K). This indicates that coupling takes place between the local nanocluster polarization and magnetization.[1] So far, these studies on 0.8PFN-0.2PMW were carried out on ceramic samples. The striking differences between PFN and PFN-PMW and the mesoscopic magnetic and electric order in PFN-PMW naturally provide impetus to study these relaxor materials in thin film form. It is also of interest to compare thin film characteristics to the results we have recently reported[7] on epitaxial single-phase films of the PFN parent compound. In Ref. 7 we have measured differences between thin films and bulk PFN, reported spin-lattice coupling in PFN thin films, and postulated the existence of superantiferromagnetic clusters as a plausible explanation for the weak ferromagnetism observed at RT in PFN. For such materials, the growth of thin film samples without impurity phases continues to be a challenging task due to compositional complexity and highly volatile Pb and Mg in these oxide compounds.[8, 9] In this letter, we report on the electric and magnetic relaxorlike behavior of 0.8PFN-0.2PMW films epitaxially grown on (001) $SrTiO_3$ (STO) single crystal substrates.

The films studied were grown by pulsed laser deposition (PLD) using a KrF excimer laser. A stoichiometric ceramic target was prepared by solid state sintering from mechanochemical synthesized 0.8PFN-0.2PMW powder.[1] The composition and structural characteristics of the films were analyzed by reflection high energy electron diffraction (RHEED) and by standard x-ray diffraction (XRD) using Cu $K\alpha$ radiation. The deposition rate was determined by modeling observed XRD Laue oscillations. To characterize the electric properties, films were grown on 50 nm $SrRuO_3$ conductive layers used as bottom electrodes and Pt was the top electrode. Macroscopic polarization-electric field (*P-E*) and local piezoresponse hysteresis loops were measured using a Sawyer-Tower circuit and a piezoforce microscope (PFM, Topometrix Explorer), respectively. The dielectric measurements were carried out using a Solartron 1260 impedance analyzer. The magnetic measurements were carried out using a Quantum Design superconducting quantum interference device magnetometer.

Figure 1(a) shows XRD $\theta$-$2\theta$ scans of 55 nm 0.8PFN-0.2PMW thin films grown at 0.2 mbar between 690 and 735 $^o$C. For films deposited at lower temperatures ($T_S \leq 700$ $^o$C), extra peaks are detected at 34.1$^o$ and 71.8$^o$, corresponding to the (*h*00)-oriented Pb-deficient cubic pyrochlore $Pb_3Nb_4O_{13}$ ($P_3N_4$, labeled $\diamond$).[8] An ordered RHEED pattern with both hexagonal spots and streaks is observed [Fig. 1(b)]. In contrast, films deposited at higher temperatures ($T_S \geq 715$ $^oC$) exhibit another ($hhh$)-oriented stoichiometric rhombohedral pyrochlore $Pb_2Nb_2O_7$ ($P_2N_2$, labeled □) or a mixture of $P_2N_2(hhh)$ and $P_3N_4(h00)$ phases. The corresponding distinctive RHEED pattern for these films is shown in Fig. 1(d). Similarly, the influence of oxygen pressure $P_{O_2}$ on phase evolution has been also explored and the optimum $P_{O_2}$ is found around 0.2 mbar. Perovskite films without any detectable impurity phases can be obtained only for a narrow deposition window ($T_S$ = 710 ± 10 $^oC$ and $P_{O_2}$ = 0.2 ± 0.1 mbar). For these 0.8PFN-0.2PMW films, only (00$l$) reflections are visible, indicative of $c$-axis orientation. The full width at half maximum (FWHM ~ 0.07$^o$) of the rocking curve for a 55 nm thick film is close to that of the substrate peak (0.065$^o$), suggesting high crystalline quality. The RHEED pattern presents fine streaks [Fig.1(c)], reflecting a smooth and well-ordered surface for these single-phase films. Furthermore, the spatial coincidence of the [100] azimuth for the film and the substrate confirms cube-on-cube epitaxial growth which is also evidenced by XRD $\phi$-scans (not shown).

Figure 2(a) illustrates the temperature dependence of the dielectric permittivity $\varepsilon'$ for a 500 nm film. Frequency dispersion of $\varepsilon'$ and a peak temperature $T_m$ is observed, indicative of dielectric relaxor behavior in 0.8PFN-0.2PMW films. The shift in $T_m$, with respect to the 250 K reported for the bulk,[1] can be due to a variation in the stoichiometry, space-charge effects[10] or strain[11] in the film. The dielectric permittivity $\varepsilon'$ is considerably broader and smaller than that in the bulk. Also the dispersion in the dielectric spectra above $T_m$ is observed in the film but not in the bulk. These peculiarities, also reported in other relaxor thin films, are generally explained by the presence of a low-permittivity passive dielectric layer at the film-substrate interface[12] and not by an intrinsic feature of the film. This interface layer contribution can be estimated[12] from the empirical scaling law $1/\varepsilon' = A + B(T-T_A)^2$ at $T > T_m$, where the fitting parameter $A$ is determined mainly by the inverse interface capacitance. Thus the intrinsic permittivity $\varepsilon'_{film}(f,T)$ of films can be reconstructed from as-measured $\varepsilon'$ using $1/\varepsilon'_{film} \approx 1/\varepsilon'-A$, as shown in the inset of Fig. 2(a). The reconstructed intrinsic permittivity of the films approaches the bulk value (~ 15000), evidencing the validity of the interface layer model.

In relaxor ferroelectrics, a modified Curie-Weiss law [13] $(\varepsilon'_{max}/\varepsilon')-1 = (T-T_m)^\gamma / 2\Delta^2$ is known to describe the relaxor permittivity empirically in the vicinity of $T_m$, and to characterize the degree of relaxation ($\gamma$) and the broadening parameter ($\Delta$). We have used this law to fit our data at 1 kHz [Fig. 2(b)]. The values extracted for $\gamma \sim 1.9$ and for $\Delta \sim$ 45 K agree with those of other ferroelectric relaxors.[10]

Relaxor dynamics, characterized by a freezing process of polar nanoregions (PNRs), is commonly analyzed by the VF relationship between the measurement frequency $f$ and $T_m$:[14]

$$f = f_0^e \exp[-E_a^e / k_B(T_m - T_{VF}^e)]$$

where $f_0^e$, $E_a^e$, and $T_{VF}^e$ are related to the attempt frequency, activation energy, and freezing temperature, respectively. A good linear fit to $\ln f \propto 1/(T_m - T_{VF}^e)$ is obtained in the frequency range of $f$ = 0.5-5 kHz [inset of Fig. 2(b)], suggesting VF-type dielectric relaxation in 0.8PFN-0.2PMW films. The fitting parameters $f_0^e$ =4.4 $\times$ 10$^7$ Hz, $E_a^e$ = 0.063 eV, and $T_{VF}^e$ ~ 270 K are physically reasonable and comparable to the bulk values.[1] Our attempts to apply the Arrhenius law to our data leads to unphysical activation energy of $E_a$ ~ 1.5 eV and attempt frequency $f_0$ ~ $10^{22}$ Hz, implying the relaxation in 0.8PFN-0.2PMW films is not Arrhenius-like.

Figure 3(a) shows temperature dependent macroscopic $P$-$E$ hysteresis loops at 1 kHz for this 500 nm film. As the temperature decreases, the lossy $P$-$E$ loop at 250 K gradually transforms to well-defined and saturated loops, owing to the reduction of leakage current. Below 200 K the remanent polarization ($P_r$) and coercive field ($E_c$) are ~ 17 $\mu$C/cm$^2$ and ~ 50 kV/cm, respectively. The wide hysteresis loop for this relaxor is attributed to the large charge disorder-induced pinning forces in 0.8PFN-0.2PMW similarly as in related cubic relaxors[15]. We also mention that the $P$-$E$ loop at RT presents an ellipse-like shape due to considerable leakage currents, which mask the true PNRs switching behavior. Therefore, we investigated the local piezoresponse hysteresis loop by PFM at RT [Fig. 3b]. It clearly reveals that the piezoresponse can be switched locally at 300 K, evidence for the existence of PNRs above $T_{VF}^e$ ~ 270 K.

Our magnetization measurements (not shown) present no evidence of an antiferromagnetic transition, consistent with the result obtained on bulk,[1] and a pronounced irreversibility between the field-cooled and zero-field-cooled (ZFC) dc magnetization starting from ~ 200 K, together with a cusp at ~ 40 K in the ZFC curve at low field (100 Oe). Such behavior is generally considered as one of the characteristic features of spin glass or cluster glass systems. We present ac magnetic susceptibility at several frequencies as a function of temperature in Fig. 4(a). The real part $\mu'$ of the complex susceptibility exhibits a marked frequency dispersion, *viz.* a systematic shift of the broad maximum with increasing frequency to higher temperature accompanied by decreasing amplitude. This is a direct manifestation of slow dynamic relaxation processes and freezing into a nonequilibrium state.[16] The origin of this magnetic relaxation behavior can be distinguished by analyzing the frequency dependence of the peak temperature $T_p$ of $\mu'(T)$. The relative shift of $T_p$ per decade of $f$, i.e., $\delta T_p = \Delta T_p/(T_p\Delta \log f$ ), is determined to be ~ 0.057, which is a typical value for spin glasses or cluster glasses (0.005-0.06), indicating the cooperative nature of the freezing process.[17] It is noteworthy that $\delta T_p$ in our case is much smaller than that obtained in superparamagnets ($\delta T_p \approx 0.1$ for these noninteracting magnetic clusters).[17] Thus, the thermal activated Arrhenius law for noninteracting magnetic clusters can be discarded in the present analysis. In contrast, the VF law, which takes into account magnetic interactions, gives a satisfactory fit with reasonable values of the relevant parameters [inset of Fig. 4(a)]: $f_0^m$ ~ $10^5$ Hz, $E_a^m$ ~ 0.007 eV, and $T_{VF}^m$ ~ 36 K. Note that the $T_{VF}^m$ and the characteristic relaxation time $\tau_0^m = 1/2\pi f_0^m$ (~$10^{-6}$ s) in films are significantly higher than those reported for bulk (~ 20 K and ~$10^{-15}$ s, respectively).[1] This is possibly due to the difference in domain sizes in the ceramics and the thin films as well as the presence of epitaxial strains in the film. Clearly further investigation is necessary to elucidate this point.

We also measured the magnetic field ($H$) dependence of the in-plane magnetization ($M$) between 5 and 300 K [Fig. 4(b)]. The slim $M$-$H$ hysteresis loops at 5 K with a coercive field $H_c \approx 360$ Oe and a remanent magnetization $M_r \approx 4$ emu/cm$^3$ also support the existence of magnetic relaxation in 0.8PFN-0.2PMW films similar to the bulk[1] and other PFN-based compounds such as 0.3PFN-0.7PMW,[18] 0.9PFN-0.1PbTiO$_3$.[19] Above $T_{VF}^m$, $H_c$ and $M_r$ decrease with increasing temperature and hysteresis persists up to 300 K [inset of Fig. 4(b)]. As conjectured in our recent work on PFN films,[7] superantiferromagnetic (SAF) cluster magnetization is plausibly at the origin for this "weak ferromagnetism" at RT, similar to PFN.[7]

In conclusion, we have grown single-phase epitaxial thin films of 0.8PFN-0.2PMW films on (001) STO using very restricted deposition parameters. We demonstrate that the films are characterized by both electric and magnetic relaxorlike behavior similar to that recently reported in ceramics. However we report significant differences in the characteristic freezing temperatures and relaxation times between thin film and bulk that suggest intrinsic features in the thin film which clearly merits further investigation.

**ACKNOWLEDGEMENTS**

This work was supported by the European Sixth Framework STREP "MULTICERAL" (Grant No. FP-6-NMP-CT-2006–032616).

**FIGURE CAPTIONS**

Fig.1 (Color online) (a) XRD $\theta$-$2\theta$ scans of films deposited at 0.2 mbar and at different temperatures. The peaks are indexed as: P: 0.8PFN-0.2PMW (reflection positions are marked by vertical dotted lines), S: STO, ◇: $P_3N_4(h00)$, □: $P_2N_2(hhh)$. (b)-(d) give corresponding RHEED patterns for three films shown in (a) with the incident beam along the STO [100] azimuth. The polygons are guides to the eye to visualize the symmetry of the diffraction pattern.

Fig.2 (Color online) (a) Temperature dependence of the as-measured dielectric permittivity $\varepsilon'$ at frequencies $f$ = 0.5 - 5 kHz. Inset: reconstructed $\varepsilon'_{film}$. (b) $\varepsilon'(T)$ at 1 kHz follows the modified Curie-Weiss law in the vicinity of $T_m$. Inset: the VF behavior of the relaxation frequency.

Fig.3 (Color online) (a) Macroscopic $P$-$E$ hysteresis loops at various temperatures for a 500 nm film. (b) Local piezoelectric hysteresis loop obtained by PFM at 300 K for a 150 nm film.

Fig.4 (Color online) (a) Temperature dependence of the real part of the ac magnetic susceptibility at various frequencies. Inset: the effective relaxation frequency follows the VF behavior. (b) Magnetic field dependence of the magnetization at 5, 50, and 300 K. Inset: the magnified central region of the hysteresis loops.

**FIGURES**

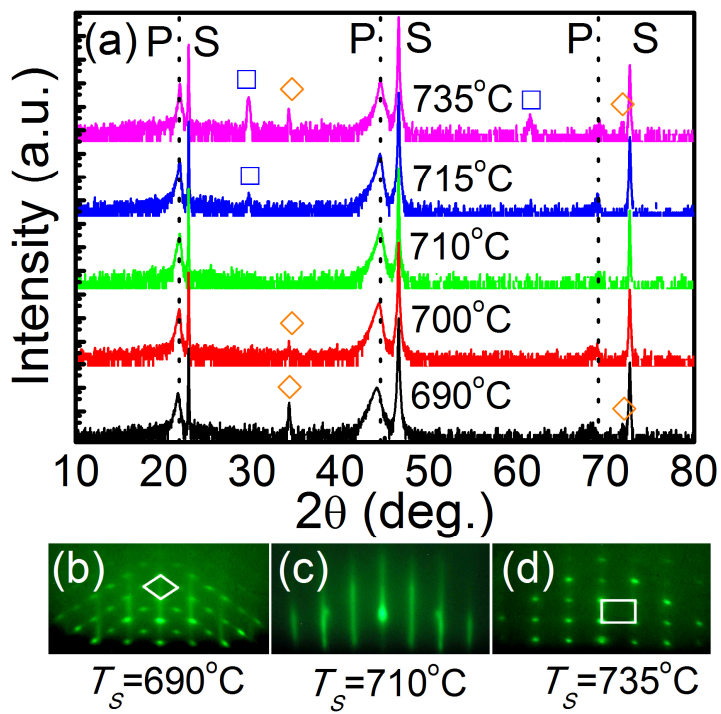

Fig.1

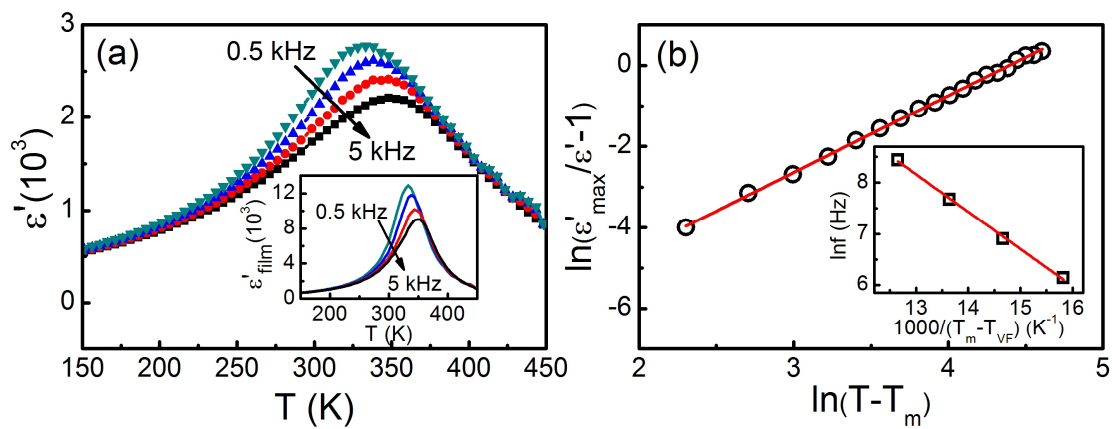

Fig. 2

0.8PFN-0.2PMW magnetoelectric relaxor thin films

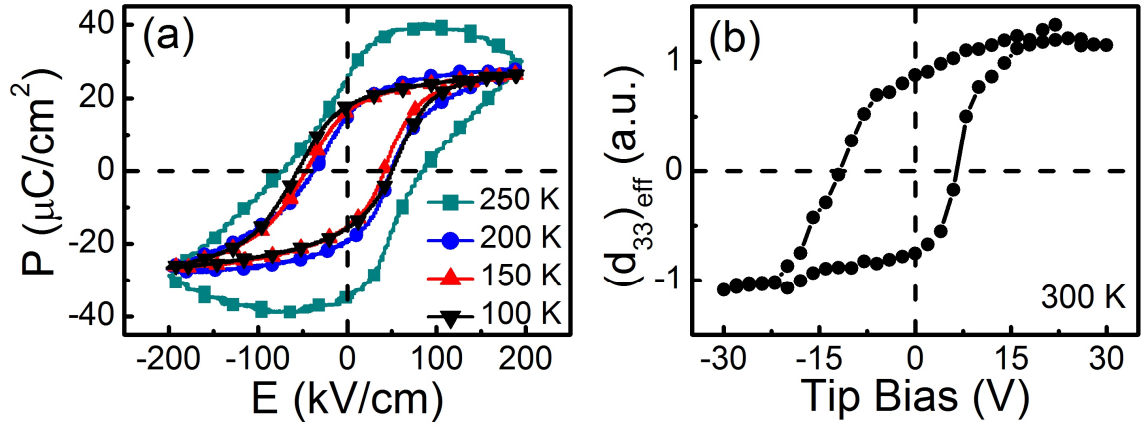

Fig. 3

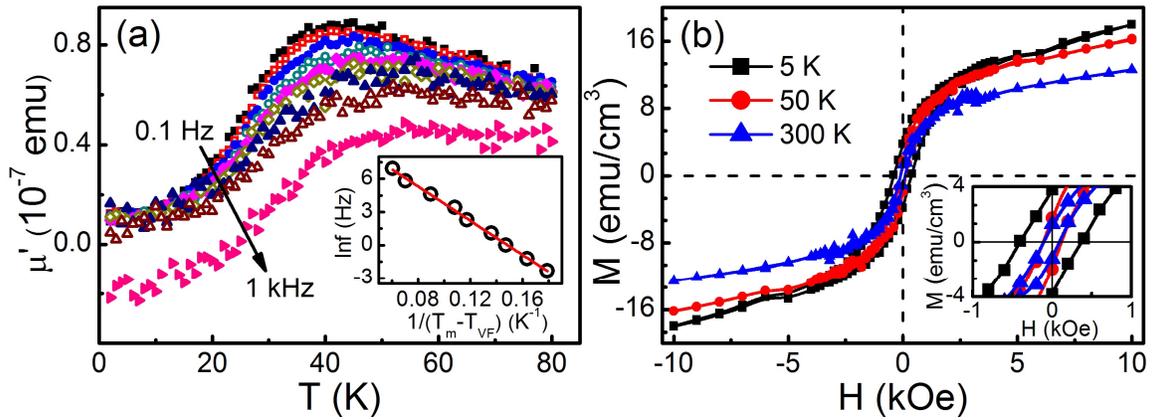

Fig. 4

0.8PFN-0.2PMW magnetoelectric relaxor thin films